\documentclass[12pt,a4paper,final]{iopart}
\usepackage{graphicx}
\begin{document}

\title[Structure, magnetism and transport in borophene nanoribbons]{How will freestanding borophene nanoribbons look like? An analysis of their possible structures, magnetism and transport properties}

Structure, magnetism and transport in borophene nanoribbons

\author{A. Garc\'{\i}a-Fuente}
\address{Departamento de F\a'{\i}sica, Universidad de Oviedo, E-33007 Oviedo, Spain}

\author{J. Carrete}
\address{LITEN, CEA-Grenoble, 17 rue des Martyrs, 38054 Grenoble Cedex 9, France}

\author{A. Vega}
\address{Departamento de F\'{\i}sica Te\'orica, At\'omica y \'Optica, Universidad de Valladolid, E-47011 Valladolid, Spain}

\author[cor1]{L. J. Gallego}
\address{Departamento de F\'{\i}sica de la Materia Condensada, Facultad de F\'{\i}sica, Universidad de Santiago de Compostela, E-15782 Santiago de Compostela, Spain}
\eads{\mailto{luisjavier.gallego@usc.es}}

\begin{abstract}
  We report a density-functional-theoretic study of the stability and electronic structure of two recently proposed borophene sheets with Pmmn
  symmetry and nonzero thickness. We then investigate nanoribbons (BNRs) derived from these nanostructures, with particular attention to
  technologically relevant properties like magnetism and electronic transport. We consider two perpendicular directions for the edges of the stripes
  as well as different lateral widths. We show that the Pmmn8 sheet, with 8 atoms in its unit cell and generated by two interpenetrating lattices, has
  a larger binding energy than the Pmmn2 sheet, with only 2 atoms per unit cell. We also use their phonon spectra to show that the mechanical
  stability of the Pmmn8 sheet is superior to that of the Pmmn2 sheet. Nanoribbons derived from Pmmn8 are not only more stable than those derived from
  Pmmn2, but also more interesting from the technological point of view. We find a rich variety of magnetic solutions, depending on the borophene
  ``mother structure'', edge orientation, width and, in the case of Pmmn8-derived BNRs, the sublattice of edge atoms. We show that one can build BNRs
  with magnetic moment in both, one or none of the edges, as well as with parallel or antiparallel magnetic coupling between the edges when magnetic;
  moreover, their electronic character can be semiconducting, metallic or half-metallic, creating a perfect spin valve at low bias.  These different
  behaviors are reflected in their densities of states, spin density and electronic transport coefficients, which are analyzed in detail. Our work
  provides a complete overview of what one may expect if nanoribbons are cut out from Pmmn sheets with a view to potential technological applications.
\end{abstract}

\vspace{2pc}
\submitto{\TDM}

\section{Introduction} \label{Sec:intro}

After the successful preparation of graphene by mechanical exfoliation of highly oriented pyrolytic graphite \cite{Novoselov} and the subsequent
investigation of its peculiar physical and chemical properties (not exhausted yet) \cite{Castroneto} growing interest has been devoted to other
single-layer or few-layer two-dimensional (2D) materials \cite{Geim}, revealing unusual properties and new phenomena that occur when charge or heat
transport are confined to a plane. Some recent examples are silicene \cite{Vogt}, molybdenum disulphide (MoS$_2$) \cite{Wang} and black phosphorene
\cite{Liu2,Shulenburger}, which are potentially useful for electronic, optoelectronic and thermoelectric applications. The number of 2D materials
discovered is increasing continuously, and it is expected that in the next few years a new generation of metallic, semimetallic, semiconductor,
superconductor, and insulator 2D materials will arise with potential technological applications.

The possibility of a boron-based analogue of graphene (borophene) was theoretically predicted some time ago, but the experimental realization of this
nanostructure has been reported very recently by Mannix {\it et al} \cite{Mannix}. Despite boron's proximity to carbon in the periodic table, its
electron deficiency prevents it from forming graphene-like planar honeycomb structures. The search for boron-based 2D structures has been complicated
and controversial. This is hardly exceptional in the history of boron compounds, full of false leads, starting with two erroneous reports of the
isolation of the element in 1808, and leading some authors to consider it as ``arguably the most complex element in the periodic table''
\cite{Oganov}. In fact, given boron's tendency towards multicenter bonding, it displays multiple bulk phases, and the formation of clusters,
fullerene-type cages and nanotubes made of boron has also also a subject of debate \cite{Kiran,Huang,Ciuparu,Oger,Szwacki,Yang,Singh,SandipDe}.

Early density functional theory (DFT) calculations suggested that single-atomic-layer boron sheets composed of triangular and hexagonal motifs were
locally stable \cite{Tang}. However, subsequent calculations predicted two novel 2D boron phases with nonzero thickness that are considerably more
stable than any of those early possibilities \cite{Zhou}. The presence of a distorted Dirac cone in the electronic band structure of one of the new phases renders
it specially interesting, as the first material with massless fermions that is not closely related to a graphene-like honeycomb structure. This new
phase of boron has a buckled structure containing eight atoms in the unit cell and belongs to the Pmmn space group; accordingly, it will be henceforth
referred to as the Pmmn8 phase, or simply the $\beta$ sheet. Recently, it has been predicted that the thermal conductivity of the $\beta$ phase of
boron is comparable to that of MoS$_2$ but shows significant in-plane anisotropy comparable to that of black phosphorene \cite{Rotational}. These
qualities may enable the efficient heat management of borophene devices in potential nanoelectronic applications.

The experimental realization of borophene has recently been reported in the form of sheets on a Ag(111) surface \cite{Mannix}. The lowest-energy
structure of the adsorbed monolayer can be constructed from distorted B$_7$ clusters using the Aufbau principle proposed by Boustani
\cite{Boustani}. Its space group symmetry is Pmmn, with lattice constants a and b equal to 2.89 \AA\ and 5.00 \AA\, respectively. However,
freestanding relaxation of this structure removes the slight corrugations along the {\bf a} direction, preserving the buckling along the {\bf b}
direction \cite{Mannix}. The resulting Pmmn structure of the freestanding borophene has two boron atoms per unit cell (henceforth it will be denoted
as the Pmmn2 or $\gamma$ phase) and lattice constants a = 2.865 \AA\ and b = 1.67 \AA. Experimental measurements, supplemented by theoretical
calculations, predict that both supported and freestanding Pmmn2 borophene are highly anisotropic 2D metals. i.e, they are conductive in the direction
of the ridges, but not across them \cite{Mannix}.

In the work described here we investigate the structural, magnetic and electronic properties of freestanding borophene nanoribbons (BNRs),
i.e. stripes of borophene with nanometer size widths. In general, the introduction of edges in a layered or non-layered 2D material has profound
effects on its properties. For instance, graphene nanoribbons (GNRs) with either zigzag or armchair edges, which have been extensively studied in the
past few years, have shown a variety of magnetic and electronic behaviors, quite different from those of the graphene sheet \cite{Castroneto, YWSon,
  Rigo, Cocchi, Longo, Martins, Martins2, Rosales, Hong, Areshkin, Hod}. Theoretical and experimental studies on nanoribbons of other 2D hexagonal
crystals, with structures similar to that of graphene, have also been performed, including those made of Si \cite{Lebegue, Ding, Cahangirov,
  Cahangirov2, DePadova, Aufray, Song, Song2, DePadova2, Song3, Kara}, Ge \cite{Lebegue, Cahangirov, Cahangirov2, Pang,Pang2}, BN \cite{Park, Barone,
  Zhang, Zheng, Zeng, Bezanilla, Wang2}, and GaN \cite{Yang2, Xiang, Li}.  However, to the best of our knowledge, calculations on the magnetic and
electronic properties of BNRs are almost nonexistent. As far as we know, the only previous work on BNRs has been Liu {\it et al}'s recent study
\cite{Liu}, who predicted that line-edge BNRs composed by a finite number of boron rows, derived from what we have called the Pmmn2 or $\gamma$ phase,
exhibit high thermal stabilities, low-resistivity Ohmic conductance and good rigidities.  In the present broader study we shall consider BNRs
conceptually cut out from both the Pmmn2 and Pmmn8 phases of boron. Our aim is twofold: i) to show that the Pmmn8 phase is indeed more stable than
the Pmmn2 phase, and ii) to show that BNRs constructed from the Pmmn8 phase are more stable and more interesting, both from a theoretical perspective
and with a view to potential technological applications, than those derived from the Pmmn2 phase. Indeed, it will be seen that, owing to the highly
anisotropic 2D structure of borophene, Pmmn8-derived BNRs exhibit a richer variety of structural, magnetic and electronic properties than those of its
other 2D analogues including graphene.

The essential technical details of the method used are sketched in section \ref{Sec:methods}, our results are presented and discussed in section 
\ref{Sec:results} and in section \ref{Sec:conclusions} we summarize our main conclusions.

\section{Details of the computational procedure} \label{Sec:methods}

We performed calculations of the structural, magnetic and electronic properties of borophene sheets and BNRs using the DFT-based code SIESTA
\cite{siesta}. The exchange-correlation potential was computed using the Perdew-Burke-Ernzerhof form of the generalized gradient approximation (GGA)
\cite{pbe}. The valence electrons of B were described with a double-$\zeta$ basis set, and their interaction with the atomic core was described by a
nonlocal, Troullier-Martins pseudopotential \cite{TM} factorized in the Kleinman-Bylander form \cite{KB}.  The B pseudopotential was generated with a
2$s^2$ 2$p^1$ electronic configuration and cutoff radii of 1.59 a.u. for both the 2$s$ and the 2$p$ orbitals.  Structures were relaxed using a
conjugated-gradient method until all forces were smaller than 0.005 eV/\AA. The periodicity of these systems was described by including 40 $k$-points
on each direction of the borophene sheets and 100 $k$-points on the growth direction of the BNRs.

To determine the phonon spectra of the borophene sheets, we employed a real-space, supercell-based method using a combination of Phonopy
\cite{Phonopy} to generate the supercell configurations and assemble the interatomic force constant matrix, and VASP \cite{vasp1,vasp2} to perform the
actual DFT calculations. We verified that our SIESTA setup afforded the same results as VASP for the borophene sheets, which gave us confidence for
employing SIESTA for the more CPU-demanding calculations on the BNRs. Since both structures under study are periodic only in two dimensions, it is
critical to ensure that force constants obtained from fulfil the requirement of rotational symmetry in order to obtain physically reasonable phonon
dispersions, for which we employed the same algorithm described in Ref. \cite{Rotational}.

\section{Results and discussion}\label{Sec:results}

\begin{figure}[ht!] 
\centering
\includegraphics[width=1.00\columnwidth]{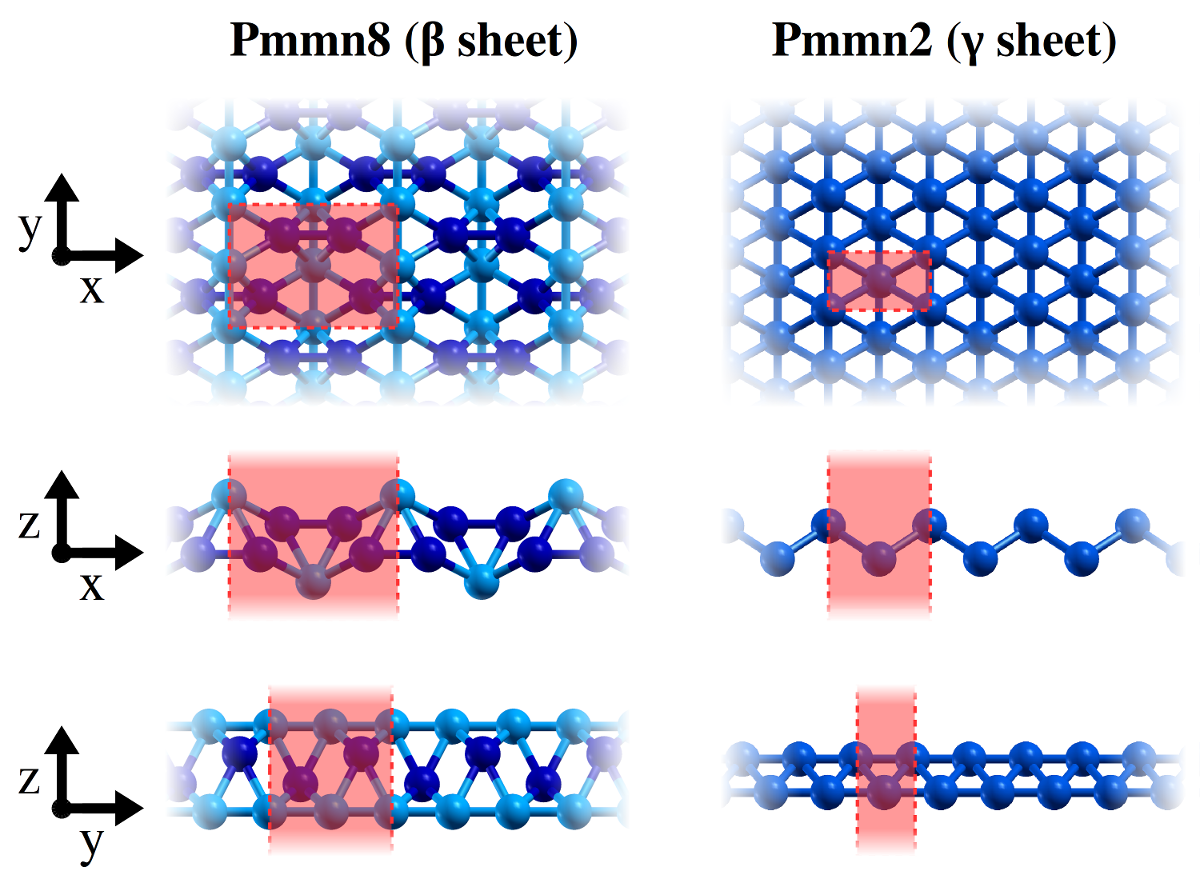}
\caption{Top, front and side views of the $\beta$ (Pmmn8) and $\gamma$ (Pmmn2) sheets of boron. The unit cell of each structure is shaded in red.
For the $\beta$ sheet, dark and light blue are used to differentiate the two types of nonequivalent atoms.}
\label{fig:periodic}
\end{figure}

\begin{figure}[ht!]
\centering
\includegraphics[width=1.00\columnwidth]{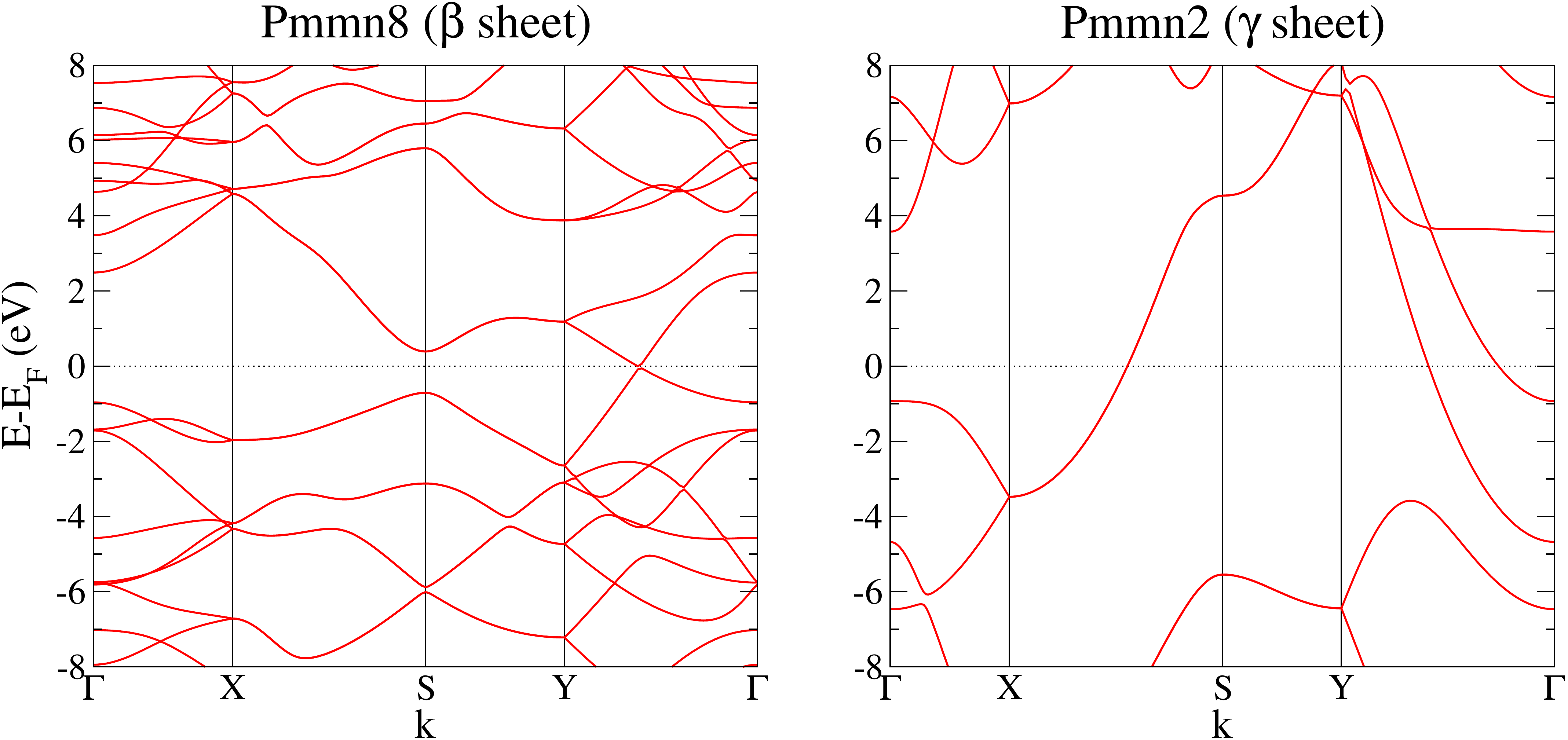}
\caption{Band structures of the $\beta$ (Pmmn8) and $\gamma$ (Pmmn2) sheets of boron.}
\label{fig:bands}
\end{figure}

As a preliminary step, we simulate the periodic $\beta$ and $\gamma$ sheets, allowing both the atomic positions and the lattice vectors to relax. The
resulting geometries and electronic band structures are shown in figure \ref{fig:periodic} and figure \ref{fig:bands}. For the $\beta$ sheet, we
obtain lattice vectors of lengths a = 4.576 \AA\ and b = 3.302 \AA, in good agreement with the results reported in Ref. \cite{Zhou}, 4.52 \AA\ and
3.26 \AA. Furthermore, the band structure of this phase presents semiconducting behavior in the $\Gamma$-X direction and a Dirac point in the
$\Gamma$-Y direction, in keeping also with the results reported in Ref. \cite{Zhou}.  The $\gamma$ sheet has lattice vectors of lengths a = 1.640 \AA\
and b = 2.934 \AA, in good agreement with the values 1.617 \AA\ and 2.865 \AA\ obtained by Mannix {\it et al} \cite{Mannix}, and its band structure
has bands crossing the Fermi level in both the $\Gamma$-Y and the X-S directions, as predicted by the same authors.

In order to determine the relative stability of the $\beta$ and $\gamma$ sheets, we computed their binding energies per atom $E_B$ using the expression 

\begin{equation}
E_B=-\frac{E_{cell}-nE_{atom}}{n},
\end{equation}

\noindent where $E_{cell}$ is the energy of a borophene unit cell, $E_{atom}$ is the energy of the isolated B atom and $n$ is the number of atoms in
the unit cell. We obtain $E_B$ = 5.843 eV for the $\beta$ sheet and 5.606 eV for the $\gamma$ sheet, indicating that the former is more stable by more than
0.2 eV per atom.

\begin{figure}[ht!] 
\centering
\includegraphics[width=1.00\columnwidth]{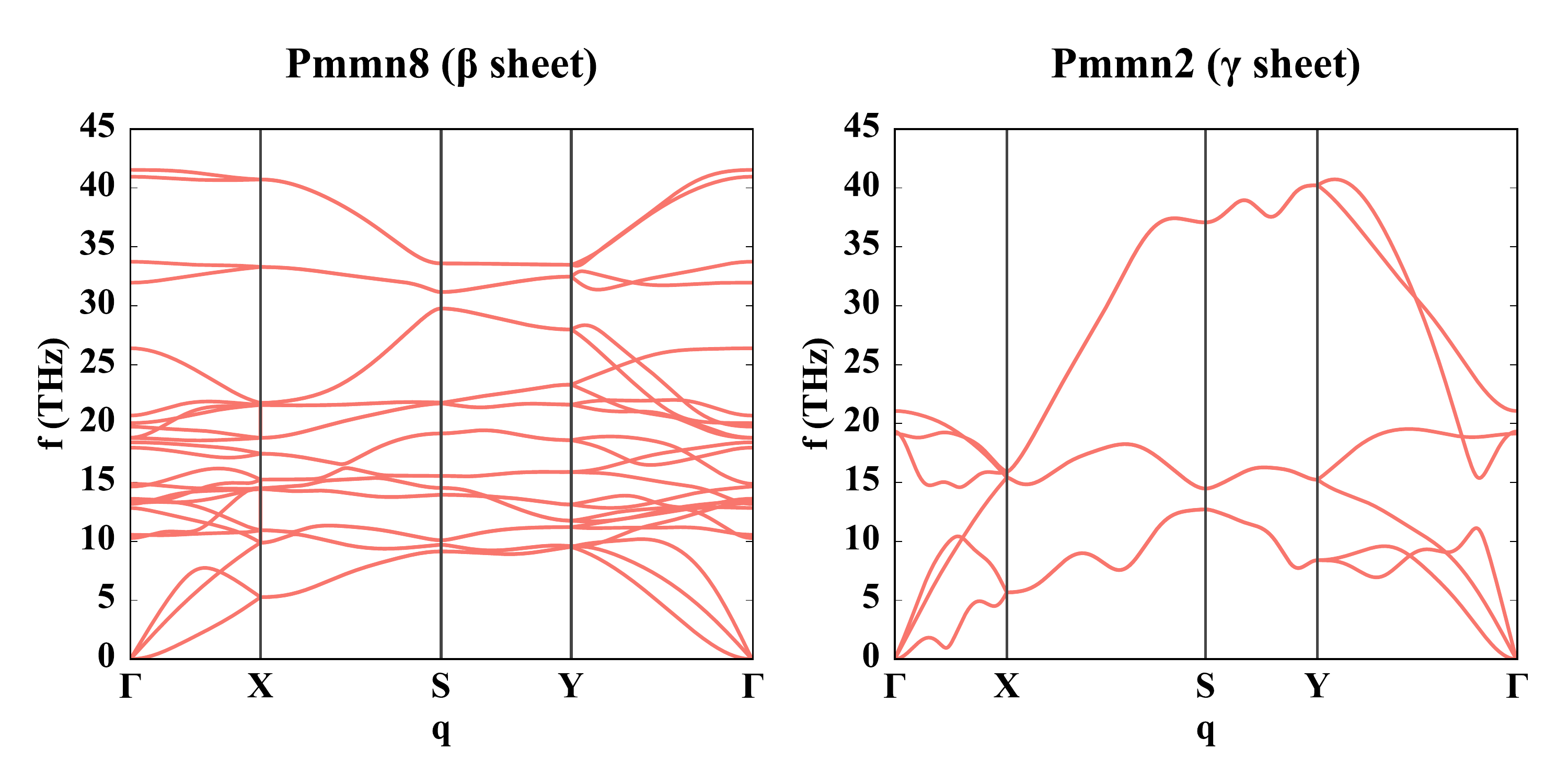}
\caption{Phonon dispersions of the $\beta$ (Pmmn8) and $\gamma$ (Pmmn2) sheets of boron.}
\label{fig:phonons}
\end{figure}

As shown in figure \ref{fig:phonons}, the phonon spectra of both the $\beta$ and $\gamma$ phases of boron consist exclusively of real frequencies,
meaning that both structures are in principle mechanically stable. In this regard, it is particularly interesting to compare our result for the phonon
spectrum of the $\gamma$ phase of boron with those included in previous studies of the same structure \cite{Mannix, SunLiWan}, Specifically, both of
those published spectra show a small region of imaginary frequencies close to the $\Gamma$ point along the $\Gamma\rightarrow X$ direction. The
spectrum from Ref. \cite{SunLiWan} also yields a finite speed of sound in the $\left|q\right| \rightarrow 0$ limit along $\Gamma\rightarrow Y$.  Both
of those features are characteristic of a lack of rotational symmetry in the force constants \cite{Rotational}. Indeed, in the case of the $\gamma$
phase of boron we have found them to be numerical artifacts introduced by the use of periodic boundary conditions in the DFT calculations, which
disappear after enforcing all the physical symmetries. Hence, our spectrum for the $\gamma$ phase does not contain the instabilities mentioned in
Refs. \cite{Mannix} and \cite{SunLiWan}, and shows a fully quadratic ZA branch along $\Gamma\rightarrow X$ and $\Gamma\rightarrow Y$. Nevertheless,
the presence of a low-frequency valley in the ZA branch between $\Gamma$ and $X$ suggests that the energetic cost of inducing a transition to the
$\beta$ ground state is modest; indeed, finite-temperature effects on the phonon spectrum could be enough to further soften those vibrational modes
and render the $\gamma$ phase unstable. The $\beta$ phase has no corresponding soft modes, and is thus a better candidate for mechanical stability in
practice.

\begin{figure}[ht!]
\centering
\includegraphics[width=1.00\columnwidth]{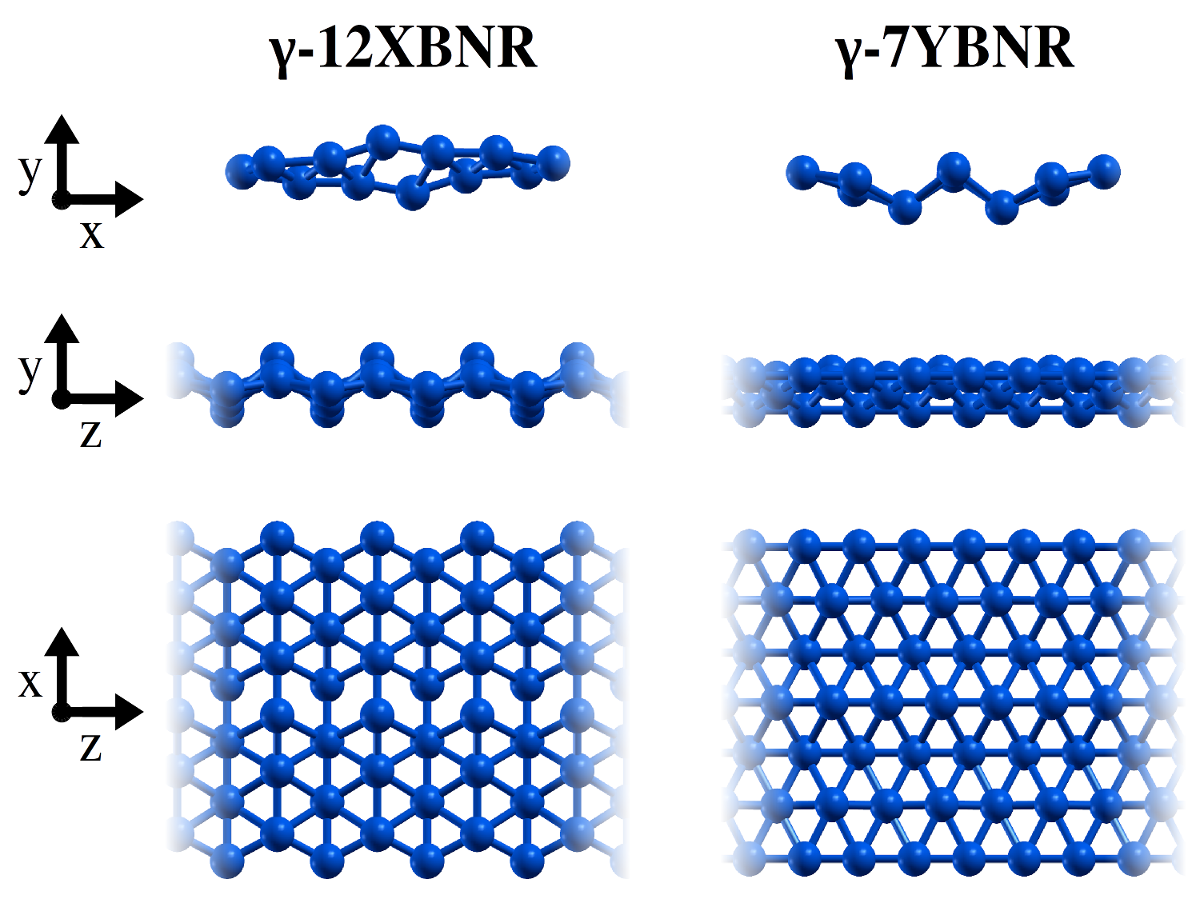}
\caption{Front, side and top views of the relaxed structures of illustrative examples of $\gamma$-NXBNRs and $\gamma$-NYBNRs.}
\label{fig:pmmn2}
\end{figure}

\begin{figure}[ht!]
\centering
\includegraphics[width=1.00\columnwidth]{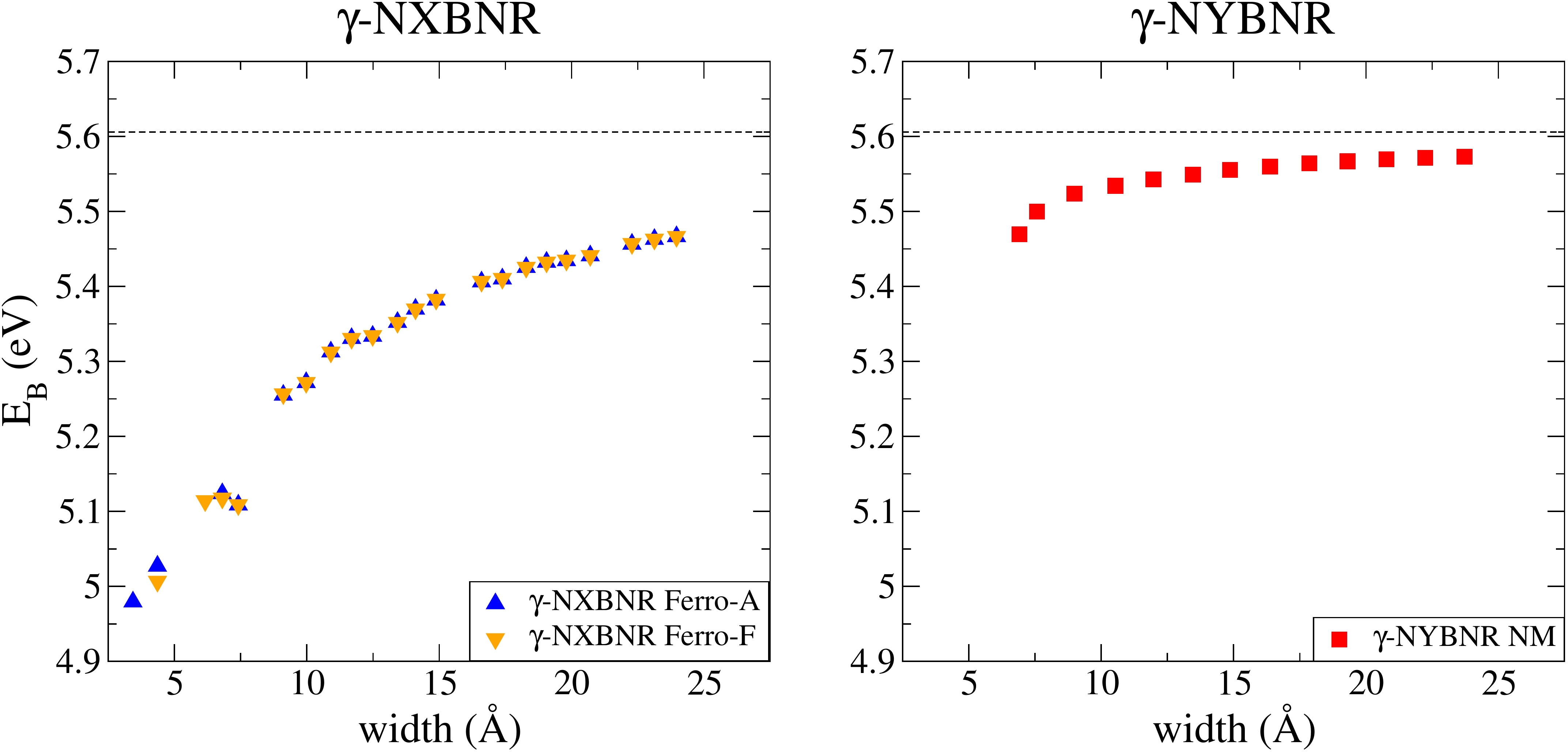}
\caption{Binding energy per atom of the $\gamma$-based BNRs with indication of their magnetic configurations, as described in the text. The dashed
  line corresponds to the binding energy for the periodic $\gamma$ sheet.}
\label{fig:bond2}
\end{figure}

In the light of the results obtained above, the best choice for investigating the properties of freestanding (or suspended) BNRs should be the $\beta$
sheet, taken as starting point to conceptually obtain quasi-one-dimensional nanostructures with differently shaped edges. This type of nanoribbons,
not explored yet, is precisely one of the main concerns of this paper. However, we have also made some calculations on BNRs derived from the less
stable $\gamma$ sheet, which are the nanostructures studied in the recent paper by Liu {\it et al} \cite{Liu}. As indicated above, our aim is to
present a broad view of the structural, magnetic and electronic properties of BNRs in general, in order to find out which are more stable and more
interesting from a fundamental point of view, as well as better candidates for potential technological applications.

In what follows we shall denote the nanoribbons derived by cutting the $\gamma$-sheet along the x and y directions as $\gamma$-NXBNRs and
$\gamma$-NYBNRs, respectively, where the first letter N is the number of B atom rows along the width of the ribbon.  In our calculations, we
considered $\gamma$-NXBNRs with N = 5-17 and $\gamma$-NYBNRs with N = 5-30, so that the maximum ribbon width is about 25 \AA. We simulated a double
unit cell in order to allow different possible magnetic arrangements within the edges as well as structural deformations, such as zigzag motifs or
Peierls-like distortions on the edges. As illustration, figure \ref{fig:pmmn2} shows the relaxed structures of $\gamma$-12XBNR and $\gamma$-7YBNR.

Figure \ref{fig:bond2} shows the binding energies and magnetic configurations obtained in our calculations for $\gamma$-NXBNRs and $\gamma$-NYBNRs.
In keeping with the results obtained by Liu {\it et al} \cite{Liu}, $\gamma$-NYBNRs are more stable than $\gamma$-NXBNRs, an effect that has been
attributed to the long-range delocalization of $\pi$ electrons along the boron rows \cite{Liu}. On the other hand, the edge atoms of $\gamma$-NYBNRs
are always non magnetic (NM), while those of $\gamma$-NXBNRs have large magnetic moments, ranging from 0.7 to 0.9 $\mu_B$. The edge atoms of
$\gamma$-NXBNRs couple ferromagnetically to each other and either ferromagnetically (Ferro-F form) or antiferromagnetically (Ferro-A form) to those of
the opposite edge. The magnetism of the edge atoms of $\gamma$-NXBNRs can be associated to their low coordination, which also leads to their low
stability.  In fact, the absence of data in figure \ref{fig:bond2} for $\gamma$-NXBNRs with N = 7,11,20 and 27 is due precisely to the strong
distortions occurring during the relaxation process of the corresponding nanoribbons, not allowing to reach stable configurations.

\begin{figure}[ht!]
\centering
\includegraphics[width=1.00\columnwidth]{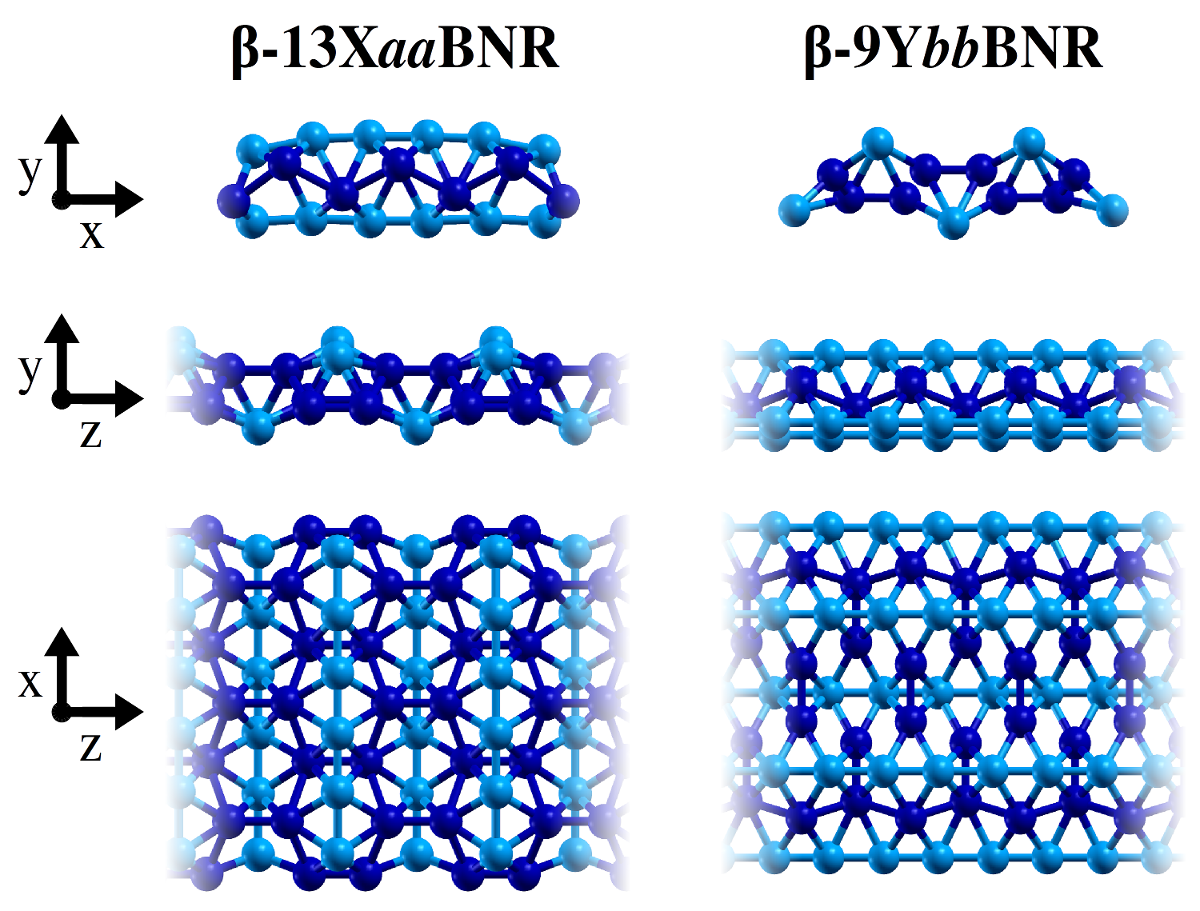}
\caption{Front, side and top views of the relaxed structures of illustrative examples of $\beta$-NX$uv$BNRs and $\beta$-NY$uv$BNRs.
Dark and light blue spheres are used to represent the two types of nonequivalent atoms (in the text, atoms of type $a$ and $b$, respectively).}
\label{fig:pmmn1}
\end{figure}

The $\beta$ sheet has two nonequivalent sets of atomic positions (or two sublattices), as illustrated in figure \ref{fig:periodic} by dark and light
blue colors. This means that, in constructing the corresponding nanoribbons, we can choose both the grow direction, x or y, and the type of B atom
($a$, dark blue, or $b$, light blue) at each edge. We represent these nanoribbons as $\beta$-NX$uv$BNR and $\beta$-NY$uv$BNR, where and $u$, $v$
$\in \{a,\,b\}$. Due to these criteria, nanoribbons with the same type of B atom at both edges ($aa$ or $bb$) have an odd value of N, while
nanoribbons with different type of B atom at each edge have an even value of N.  We simulated $\beta$-NX$uv$BNR with N = 6-30 and $\beta$-NY$uv$BNR
with N = 5-27, so that the maximum width for these nanoribbons is close to 25 \AA.  Illustrative examples of the relaxed structures of some of these
types of nanoribbons are given in figure \ref{fig:pmmn1}.

\begin{figure}[ht!]
\centering
\includegraphics[width=1.00\columnwidth]{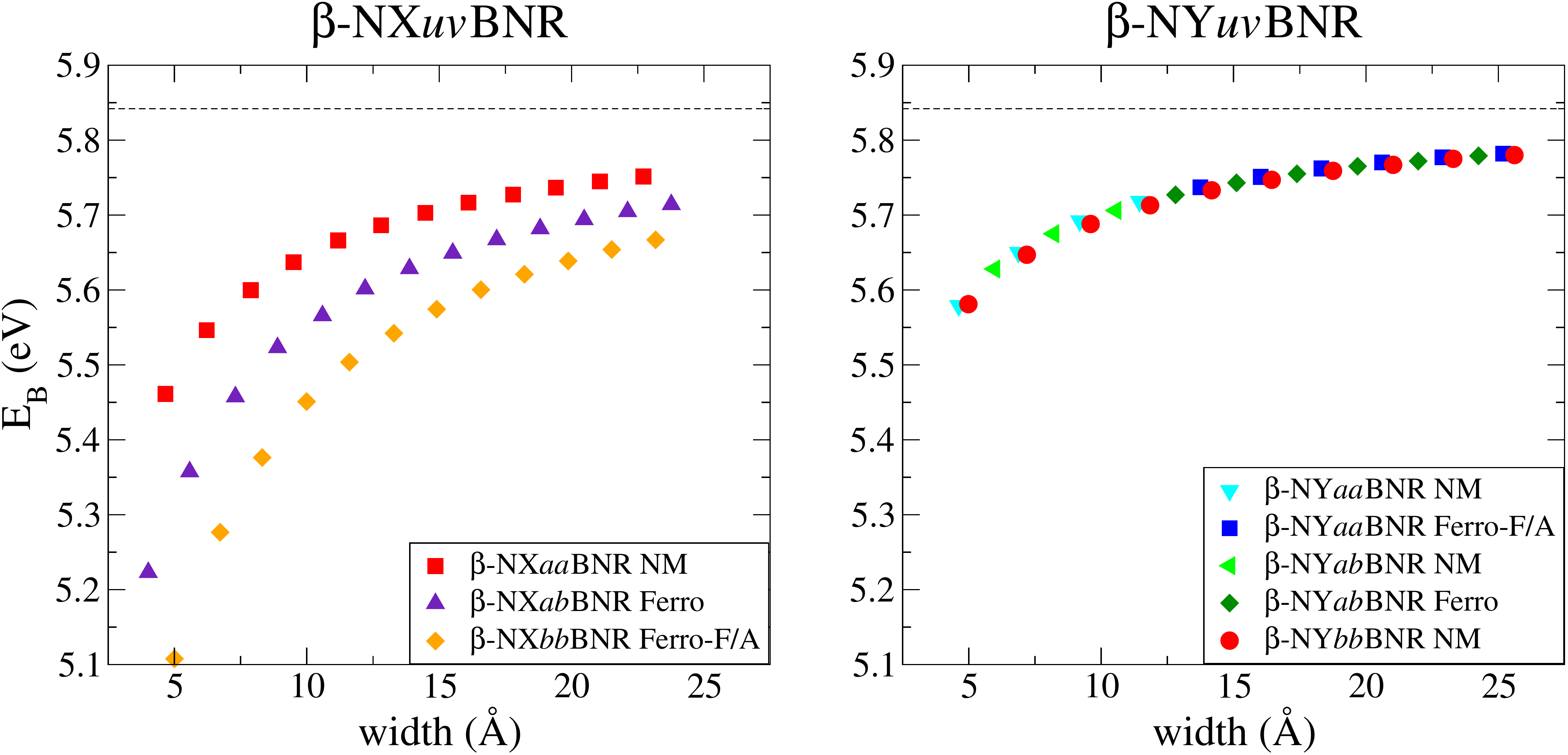}
\caption{Binding energy per atom of the $\beta$-based BNRs with indication of their magnetic configurations, as described in the text. The dashed line
  corresponds to the binding energy for the periodic $\beta$ sheet.}
\label{fig:bond1}
\end{figure}

\begin{figure*}[ht!]
\centering
\includegraphics[width=1.00\textwidth]{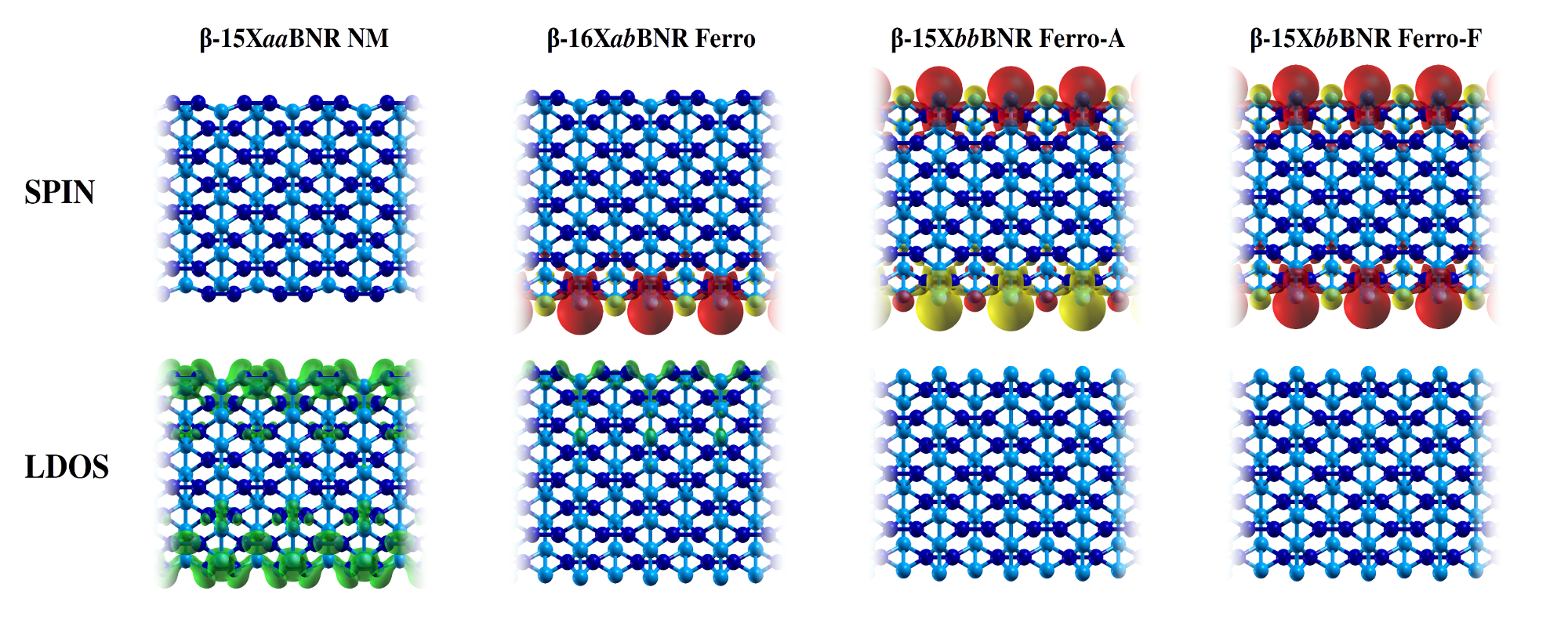}
\caption{Total spin density (upper panels) and local density of states around the Fermi level (lower panels) for several $\beta$-NX$uv$BNR, with
  indication of their magnetic configuration (see text). Red (yellow) color is used for the positive (negative) net value of the spin density; green
  color to represent the total density of states, sum of the two spin components.}
\label{fig:Xrholdos}
\end{figure*}

\begin{figure}[ht!]
\centering
\includegraphics[width=1.00\columnwidth]{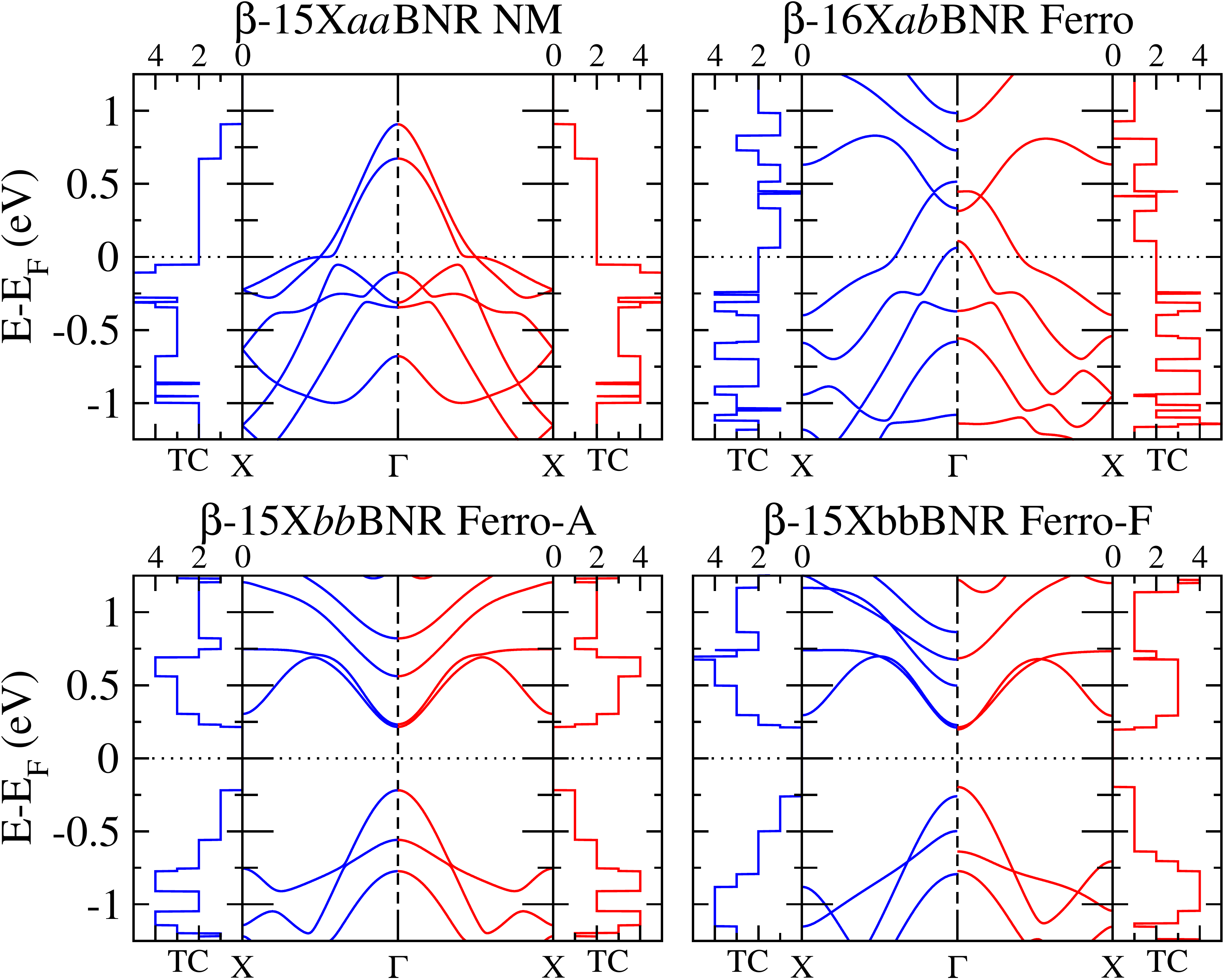}
\caption{Band structure and transmission channels (TC) for several $\beta$-NX$uv$BNR, with indication of their magnetic configuration (see text). Red
  and blue colors are used for the majority and minority spin components, respectively.}
\label{fig:xbands}
\end{figure}

The binding energies and magnetic configurations predicted in our calculations for $\beta$-NX$uv$BNR and $\beta$-NY$uv$BNR are shown in figure
\ref{fig:bond1}. Note that nanoribbons derived from the $\beta$ sheet are more stable than those derived from the $\gamma$ sheet, thus keeping the
relative stability of the original two sheets. On the other hand, the more complex structure of the $\beta$ sheet yields nanoribbons with a larger
number of possible configurations. For $\beta$-NX$uv$BNRs, there is an energetic preference for $a$-type edges, i.e., $\beta$-NX$aa$BNRs are the most
stable ribbons, followed by $\beta$-NX$ab$BNRs and $\beta$-NX$bb$BNRs. Figure \ref{fig:Xrholdos} shows the total spin density and the spatial local
density of states (LDOS) around the Fermi level of a representative sample of $\beta$-NX$uv$BNRs, and figure \ref{fig:xbands} shows their
corresponding band structures and transmission channels. We use the number of transmission channels available at a given energy, i.e., the number of
electronic bands containing states with this energy, as a estimate of the quasi-ballistic transmission for low temperature and low bias in the
Landauer formalism \cite{Landauer}.  We find that $a$-type edge atoms are NM, while $b$-type edge atoms are magnetic, with alternative atoms of that
edge having a strong magnetic moment close to 0.92 $\mu_B$ coupled antiferromagnetically to their neighbouring edge atoms, which have a small magnetic
moment close to 0.05 $\mu_B$ (see figure \ref{fig:Xrholdos}). Since the coupling between unit cells is ferromagnetic, $\beta$-NX$aa$BNRs have only NM
configurations, $\beta$-NX$ab$BNRs have configurations which are magnetic only at one edge (denoted Ferro in figures
\ref{fig:bond1}-\ref{fig:xbands}), and $\beta$-NX$bb$BNRs have two possible configurations, Ferro-F and Ferro-A, depending on the magnetic coupling
(parallel or antiparallel) between the edges. The Ferro-F and Ferro-A configurations of $\beta$-NX$bb$BNRs are degenerate down to meV precision,
indicating a very weak exchange coupling along the width of the ribbon (they are represented in figure \ref{fig:bond1} as Ferro-F/A).

The band structures of figure \ref{fig:xbands} show that the type of edge atom has drastic effects on the transmission character of the
$\beta$-NX$uv$BNRs.  Specifically, states associated to transmission channels close to the Fermi level are only localized for ribbons with edge atoms
of type $a$.  This makes both $\beta$-NX$aa$BNRs and $\beta$-NX$ab$BNRs metallic. Interestingly, although $\beta$-NX$ab$BNRs are ferromagnetic at one
edge, there are not important differences between the transmission channels for both spin components, as transmission is driven by the nonmagnetic
edge.  On the other hand, both Ferro-A and Ferro-F $\beta$-NX$bb$BNRs are semiconducting with a similar band gap of around 0.5 eV.  This contrasts
with the behaviour of hydrogen-passivated zigzag GNRs, in which the Ferro-F configuration is metallic \cite{Rigo, Martins}. The result obtained for
Ferro-F $\beta$-NX$bb$BNRs is counterintuitive since one would expect the Ferro-F solution to have metallic behavior due to the partial filling
(emptying) of a majority (minority) spin band. However, what we found is that one band is completely filled (unfilled), leading to an integer value
of 2 $\mu_B$/unit cell for the magnetic moment of these ribbons.

\begin{figure*}[ht!]
\centering
\includegraphics[width=1.00\textwidth]{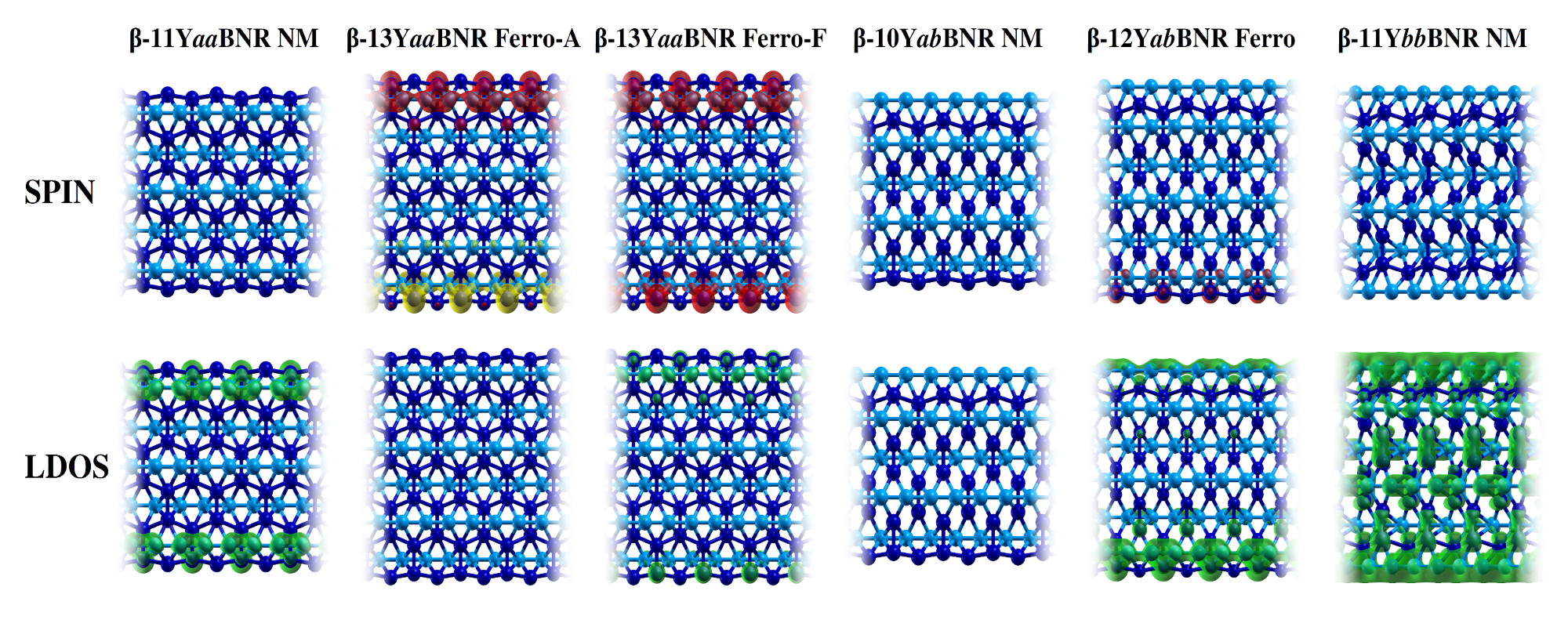}
\caption{Total spin density (upper panels) and local density of states around the Fermi level (lower panels) for several $\beta$-NY$uv$BNR, with
  indication of their magnetic configuration (see text). Red (yellow) color is used for the positive (negative) net value of the spin density; green
  color to represent the total density of states, sum of the two spin components.}
\label{fig:Yrholdos}
\end{figure*}

\begin{figure}[ht!]
\centering
\includegraphics[width=1.00\columnwidth]{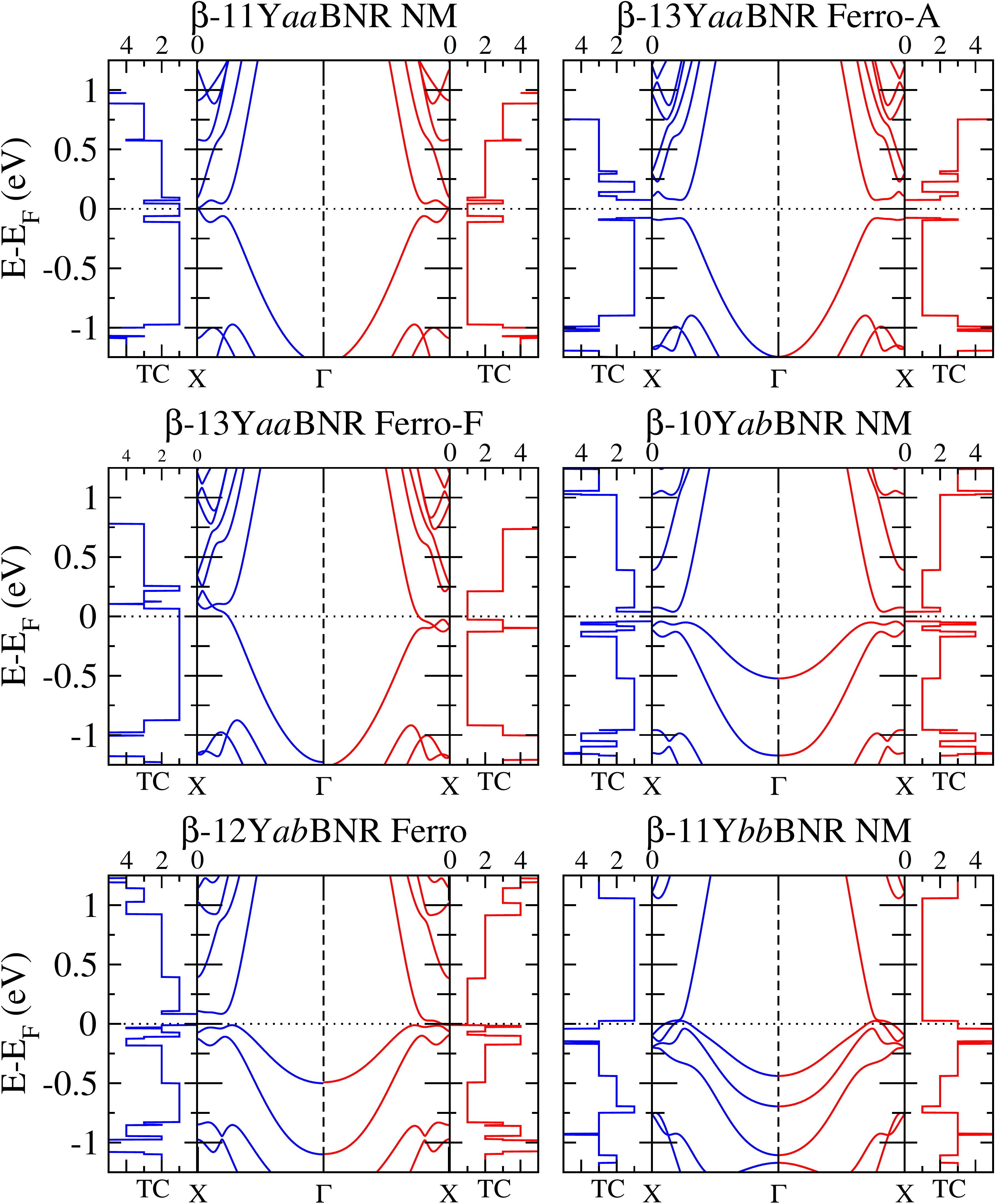}
\caption{Band structure and transmission channels (TC) for several $\beta$-NY$uv$BNR, with indication of  their magnetic configuration (see text).
Red and blue colors are used for the majority and minority spin components, respectively.}
\label{fig:ybands}
\end{figure}

The $\beta$-NY$uv$BNRs are the most stable of all the nanoribbons investigated. Their binding energies depends mainly on the width of the ribbon,
irrespectively of their structure and magnetic configuration (figure \ref{fig:bond1}). The edge atoms of type $b$ of these nanoribbons are NM, and
those of type $a$ are NM for very short widths but become magnetic for widths larger than around 12 \AA. Since the magnetic coupling along the edges
for the magnetic configurations is always ferromagnetic, $\beta$-NY$aa$BNRs are either NM or Ferro-F/Ferro-A depending on their widths,
$\beta$-NY$ab$BNRs are NM or ferromagnetic at one edge depending also on their widths, and $\beta$-NY$bb$BNRs are always NM.  Figure
\ref{fig:Yrholdos} shows the total spin density and the spatial LDOS around the Fermi level of a representative sample of $\beta$-NY$uv$BNRs, and
figure \ref{fig:ybands} shows their band structures and transmission channels.  The magnetizations of magnetic $\beta$-NY$uv$BNRs are smaller than
those of magnetic $\beta$-NX$uv$BNRs.  For magnetic $\beta$-NY$uv$BNRs, $a$-type edges have a magnetic moment close to 0.10 $\mu_B$ at one of the two
atoms of the edge for each unit cell, and a magnetic moment of around 0.05 $\mu_B$ located at each of the $b$-type neighbors of that atom.  The
Ferro-A configurations of $\beta$-NY$aa$BNRs are always more stable than the Ferro-F configurations, but their binding energies per atom differ in
less than 1 meV (they are represented together in figure \ref{fig:bond1}).

The results shown in figures \ref{fig:Yrholdos} and \ref{fig:ybands} allow pointing out some similarities between $\beta$-NY$uv$BNRs and (hydrogen
passivated) zigzag GNRs \cite{Rigo, Martins}. Specifically, $\beta$-NY$aa$BNRs present wide $\pi$-like bonding and antibonding bands around the Fermi
level.  For narrow $\beta$-NY$aa$BNRs, the magnetization is not favored and the ribbons are metallic due to the contact between bonding and
antibonding edge states. However, for wider nanoribbons the magnetization of the edge states lead to the semiconducting (Ferro-A) and metallic
(Ferro-F) possibilities which are found in zigzag GNRs \cite{Rigo, Martins}.  It should be noted that narrow $\beta$-NY$ab$BNRs are semiconducting,
and wider $B$-NY$ab$BNRs are metallic just for the majority spin component, thus becoming a perfect spin valve. Although the spin polarization is
mostly located at the $a$-type edge, the transmission is driven by states located at both edges.  A similar behavior has been observed for zigzag GNRs
in which the magnetism at one edge is quenched by the adsorption of a molybdenum chain \cite{mognr}. Finally, $\beta$-NY$bb$BNRs are metallic, with
several transmission channels at the Fermi level localized not only at the edges, but through all the width of the ribbon.

\section{Summary and conclusions}\label{Sec:conclusions}

In this work we present a DFT study of two recently proposed 2D phases of boron \cite{Mannix,Zhou}, that we call the $\beta$ and $\gamma$ phases, and of their
quasi-one-dimensional derivatives $\beta$-NX$uv$BNRs, $\beta$-NY$uv$BNRs, $\gamma$-NXBNRs and $\gamma$-NYBNRs. The computed phonon spectra of the
$\beta$ and $\gamma$ phases show that both of them are, in principle, mechanically stable. However, a low-frequency valley in the ZA branch of the
spectrum of the $\gamma$ phase suggests that thermal effects could render it unstable.

Regarding the boron nanoribbons, our results support the recent finding \cite{Liu} that $\gamma$-NYBNRs, having non-magnetic linear edges, are more
stable than $\gamma$-NXBNRs, which have less coordinated, zigzag edge atoms with magnetic moments. However, $\beta$-NX$uv$BNRs and $\beta$-NY$uv$BNRs
are noticeably more stable regardless of the lateral width, particularly the latter that can expose two different kinds of atomic arrangements at each
edge, and exhibit a rich variety of magnetic solutions. Specifically, in these nanoribbons, more coordinated edge atoms (of type $b$) are NM, while
less coordinated edge atoms (of type $a$) are NM for very short widths but become magnetic for widths larger than around 12 \AA. Since the magnetic
coupling along the edges for the magnetic configurations is always ferromagnetic, $\beta$-NY$aa$BNRs are either NM or Ferro-F/Ferro-A depending on
their widths, $\beta$-NY$ab$BNRs are NM or ferromagnetic at one edge depending also on their widths, and $\beta$-NY$bb$BNRs are always NM. Moreover,
the binding energy of the $\beta$-NY$uv$BNRs depends mainly on the width of the ribbon, irrespectively of their structure and magnetic configuration,
so that these nanoribbons present structural and magnetic multistability, a fact that could be advantageous in nanotechnology; for instance, one could
use an external field to access to a different magnetic state. The different magnetic states of the nanoribbons are related to their electronic
structure. Specifically, narrow $\beta$-NY$aa$BNRs are metallic, while for wider $\beta$-NY$aa$BNRs the magnetization of the edges lead to
semiconducting (Ferro-A) or metallic (Ferro-F) possibilities. Narrow $\beta$-NY$ab$BNRs are semiconducting, and wider $\beta$-NY$ab$BNRs are metallic
just for the majority spin component, thus becoming a perfect spin valve at low bias. Finally, $\beta$-NY$bb$BNRs are metallic, with several
transmission channels at the Fermi level localized through all the width of the nanoribbon.

Overall, our results show that one can build BNRs of different structure and with magnetic moment in both, one or none of the edges, as well as with
parallel or antiparallel magnetic coupling between the edges when magnetic; and with semiconducting, metallic or half-metallic character, thus
producing a perfect spin valve at low bias. It is these features that can be taken into account in the future for the design of devices involving
these boron nanostructures.

\section*{Acknowledgments}

This work was supported by the Spanish Ministry of Economy and Competitiveness (Projects FIS2012-33126 and FIS2014-59279-P) and by the Xunta de
Galicia (AGRUP2015/11), in conjunction with the European Regional Development Fund (FEDER).

\section*{References}

\bibliography{bibliography}

\end{document}